\documentclass[a4paper, 11pt]{article}
\usepackage{latexsym}
\usepackage{amssymb}
\usepackage{amsmath}
\usepackage{amsfonts}
\usepackage{bbm}
\usepackage{graphicx}
\usepackage{float}
\usepackage[english]{babel}
\usepackage{multirow}
\usepackage{float}
\usepackage{soul,color}
\restylefloat{table}
\usepackage{caption}
\usepackage{placeins}
\usepackage{hyperref}
\usepackage{wrapfig}
\usepackage{cleveref}
\usepackage{enumitem}
\usepackage[toc,page]{appendix}
\usepackage[normalem]{ulem}
\useunder{\uline}{\ul}{}
\usepackage{ltxtable}
\usepackage{subfig}
\usepackage{todonotes}
\setuptodonotes{inline}
\usepackage{longtable}

\newcolumntype{L}[1]{>{\raggedright\let\newline\\\arraybackslash\hspace{0pt}}m{#1}}
\newcolumntype{C}[1]{>{\centering\let\newline\\\arraybackslash\hspace{0pt}}m{#1}}

\makeatletter
\def\@maketitle{
  \begin{center}
    {\Huge \bfseries \sffamily \@title }\\[3ex]
    {\Large  \@author}\\[2ex]
    \includegraphics[width = 30mm]{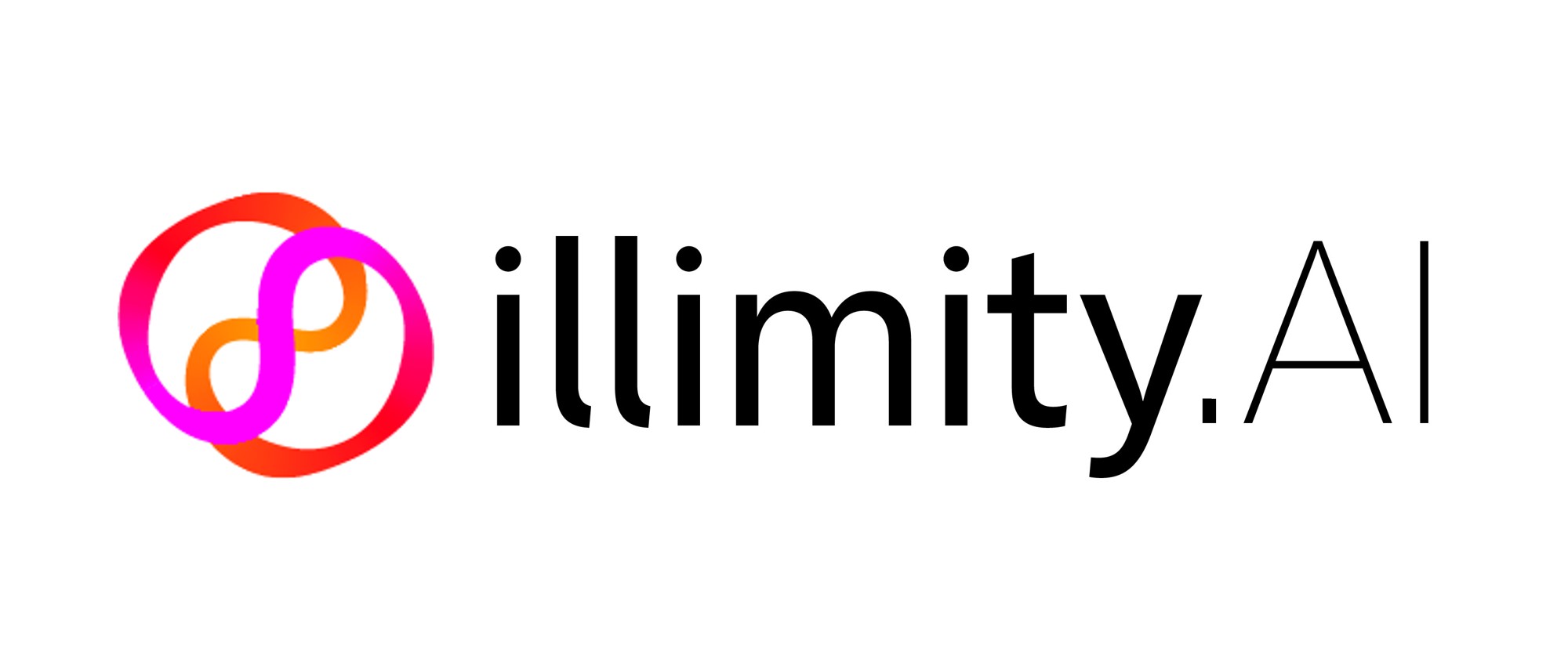}\\[4ex]
    \@date\\[4ex]
\end{center}}
\makeatother
\author{O. Didkovskyi\thanks{Equal contribution} A. Vidali\footnotemark[1], N. Jean\thanks{corresponding author: nicola.jean@illimity.com}, and G. Le Pera}
\title{Temporal-Aligned Meta-Learning for Risk Management: A Stacking Approach for Multi-Source Credit Scoring}

\begin{document}\maketitle

\begin{center}
\end{center}

\vspace{10mm}

\begin{abstract}
  \noindent
  This paper presents a meta-learning framework for credit risk assessment of Italian Small and Medium Enterprises (SMEs) that explicitly addresses the temporal misalignment of credit scoring models.
  The approach aligns financial statement reference dates with evaluation dates, mitigating bias arising from publication delays and asynchronous data sources.
  It is based on a two-step temporal decomposition that at first estimates annual probabilities of default (PDs) anchored to balance-sheet reference dates (December 31st) through a static model. Then it models the monthly evolution of PDs using higher-frequency behavioral data.
  Finally, we employ stacking-based architecture to aggregate multiple scoring systems, each capturing complementary aspects of default risk, into a unified predictive model.
  In this way, first level model outputs are treated as learned representations that encode non-linear relationships in financial and behavioral indicators, allowing integration of new expert-based features without retraining base models.

  This design provides a coherent and interpretable solution to challenges typical of low-default environments, including heterogeneous default definitions and reporting delays.
  Empirical validation shows that the framework effectively captures credit risk evolution over time, improving temporal consistency and predictive stability relative to standard ensemble methods.

\end{abstract}

\bigskip
{\bf JEL} Classification codes: C45, C55, G24
{\bf AMS} Classification codes: 91G40
\bigskip


\newpage

\tableofcontents

\newpage


\section{Introduction}
\label{section:intro}

Credit risk assessment remains a core challenge for financial institutions, requiring the synthesis of multiple information sources that describe different facets of a borrower’s financial health \cite{mcneil2005quantitative}.
Modern credit scoring must capture non-linear relationships between indicators \cite{addo2018credit,khandani2010consumer,bancaditalia2026misp73}, integrate heterogeneous data (financial statements, bureau records, expert indicators, macro variables),
and respect the temporal characteristics and reporting delays inherent to each source.

A specific but widespread operational problem is the temporal misalignment:
data used for scoring are often referenced to different dates than the evaluation (application or monitoring) date.
For example, annual balance sheets dated December 31 may only become available six months later; therefore, a model trained to predict default within one year of the balance-sheet date is, at evaluation time, predicting over a horizon that has already been partially observed.
This mismatch systematically biases risk estimates and is particularly consequential for Small and Medium Enterprises (SMEs), where information asymmetry and sparse reporting amplify sensitivity to stale data \cite{gupta2015forecasting}.
Existing multi-source fusion and temporal models handle feature fusion but typically assume aligned reference points and thus do not directly address this reference-to-evaluation date gap \cite{chen2025credit,quan2024credit,wang2018,lim2021}.

We approach this problem by combining two complementary base models that capture different risk perspectives:
\begin{itemize}
  \item \textbf{CRD Model} \cite{provenzano2020machine}: Leverages balance sheet data, providing comprehensive financial indicators but updated annually with 3-9 month publication lags.
  \item \textbf{Behavioral Model (BHV)} \cite{didkovskyi2024cross}: Utilizes Central Credit Register data, capturing borrower payment patterns and credit utilization with monthly updates but lacking the granularity of accounting statements.
\end{itemize}

These models display complementary asymmetries: CRD is deep but stale; BHV is timely but incomplete, yet neither alone resolves the temporal alignment problem.

Thus we propose a two-step framework that explicitly models the temporal gap between data reference dates and evaluation dates.
First, a \textbf{static model} is created to estimate annual PD anchored to December 31st using balance sheet data.
After, a \textbf{dynamic model} is trained on outputs of the static model to captures monthly PD evolution by integrating behavioral updates through temporal aggregation and meta-learning.
This approach yields PIT-consistent scores for both origination and monitoring and avoids full retraining when adding new indicators.

Our main contributions are:
\begin{enumerate}
  \item A temporal decomposition that separates static annual assessment from dynamic monthly evolution and explicitly tackles temporal misalignment between data sources.
  \item A point-in-time consistency methodology for aligning multi-frequency sources to common reference dates, producing PIT-adjusted PD estimates.
  \item A modular meta-learning architecture that integrates base-model embeddings and allows new indicators to be added without revalidating underlying models.
\end{enumerate}

The remainder of the paper is organized as follows. Section 2 describes the dataset and data-availability constraints. Section 3 defines targets and preprocessing. Section 4 details the modeling framework, including the static anchor model, dynamic update mechanism, and the meta-learner. Section 5 presents empirical results and validation metrics. Section 6 discusses limitations, operational considerations, and future work.

\section{Dataset} 
\label{section:data}

Our analysis utilizes multiple data sources covering Italian SMEs, with the focus to the companies within the illimity bank's portfolio, including both direct debtors and participants in joint ventures.
These sources are (i) Balance sheet data of the illimity debtors, (ii) Central Credit Register Data (CR) \cite{cr1991},
(iii) Expert-Based Indicators that may be created from both internal and external data.
CR data represents the bottleneck in our framework as it is only available for companies within the internal portfolio. 


\begin{table}[h]
  \centering
  \caption{Dataset Overview}
  \begin{tabular}{lcccc}
    \hline
    \textbf{Data Source} & \textbf{Period} & \textbf{Update Frequency} & \textbf{Sample Size} & \textbf{Lag} \\
    \hline
    Balance Sheets & 2017-2023 & Annual & 18,454 & 3-9 months \\
    Central Credit Register & 2017-2024 & Monthly & 18,454 & 2 months \\
    Expert Indicators & 2017-2024 & Variable & 18,454 & 2 months \\
    \hline
  \end{tabular}
  \label{tab:dataset_structure}
\end{table}


\subsection{On data processing}
\label{subsection:data_processing}

\begin{figure}[ht]
  \centering
  \includegraphics[width=0.85\textwidth]{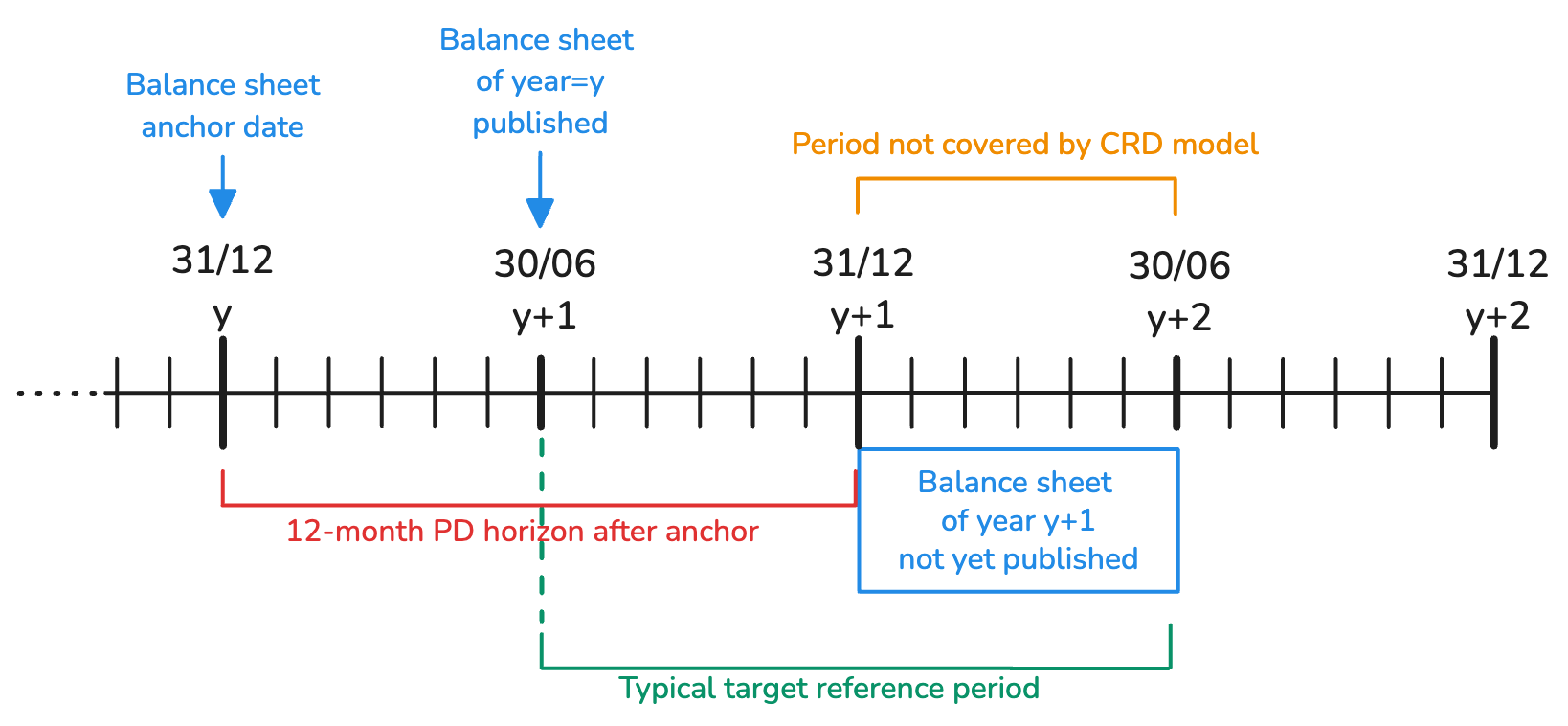}
  \caption{Temporal misalignment: December 31st balance sheets published 6 months later create a prediction gap with the 12-month horizon}
  \label{fig:temporal_misalignment}
\end{figure}

A fundamental challenge arises from the time required to process information by external agencies.
Balance sheets for December 31st become available 3-9 months later;
CR data has a 2-month lag; internal data provides near real-time updates.
Figure \ref{fig:temporal_misalignment} illustrates how this creates systematic temporal misalignment where the CRD model's 12-month prediction horizon extends beyond target period.

\subsection{Point-in-Time Consistency}
\label{subsection:point_in_time_consistency}

To address temporal misalignment we introduce point-in-time consistency:
identifying the latest information related to the reference period being modeled.
It ensures that the predictions are based on data that characterize the situation at the decision point.

Given that balance sheets represent the situation as of December 31st, the proposed consistency describes an ideal scenario wherein for each December 31st,
the model estimates probability of default for the next year using CR data aligned with this date.

\subsection{Features}
\label{subsection:features}

The meta-learner's base features are standardized logit-transformed PDs from established models:
\begin{itemize}
  \item $logit(PD_{bhv})^*$ - standardized behavioral score
  \item $logit(PD_{crd})^*$ - standardized CRD score
\end{itemize}


where the logit transformation and standardization is applied to ensure consistent scaling across different probability ranges.

For dynamic model we adopt exponentially weighted moving average of $PD_{bhv}$ to account for dynamics of monthly behavioral PDs.
The parameter $\alpha$ is calibrated on internal data.

\section{Target}
\label{subsection:target}

We indicate default event as one of the following indicators: default rettificato (Bank of Italy system-level definition), Orbis bankruptcy target, and the internal default flag.
Then manual validation corrects classification inconsistencies: companies in bankruptcy or liquidation proceedings confirmed through manual inspection are added as defaults, while false positives are removed


\subsection{Static Model Target}
\label{subsection:static_target}

The static target identifies at least one month within the next 12 months where the default occurs, starting from December 31st (balance sheet reference dates).

\subsection{Dynamic Model Target}
\label{subsection:dynamic_target}

\begin{figure}[ht!]
  \centering
  \includegraphics[width=0.85\textwidth]{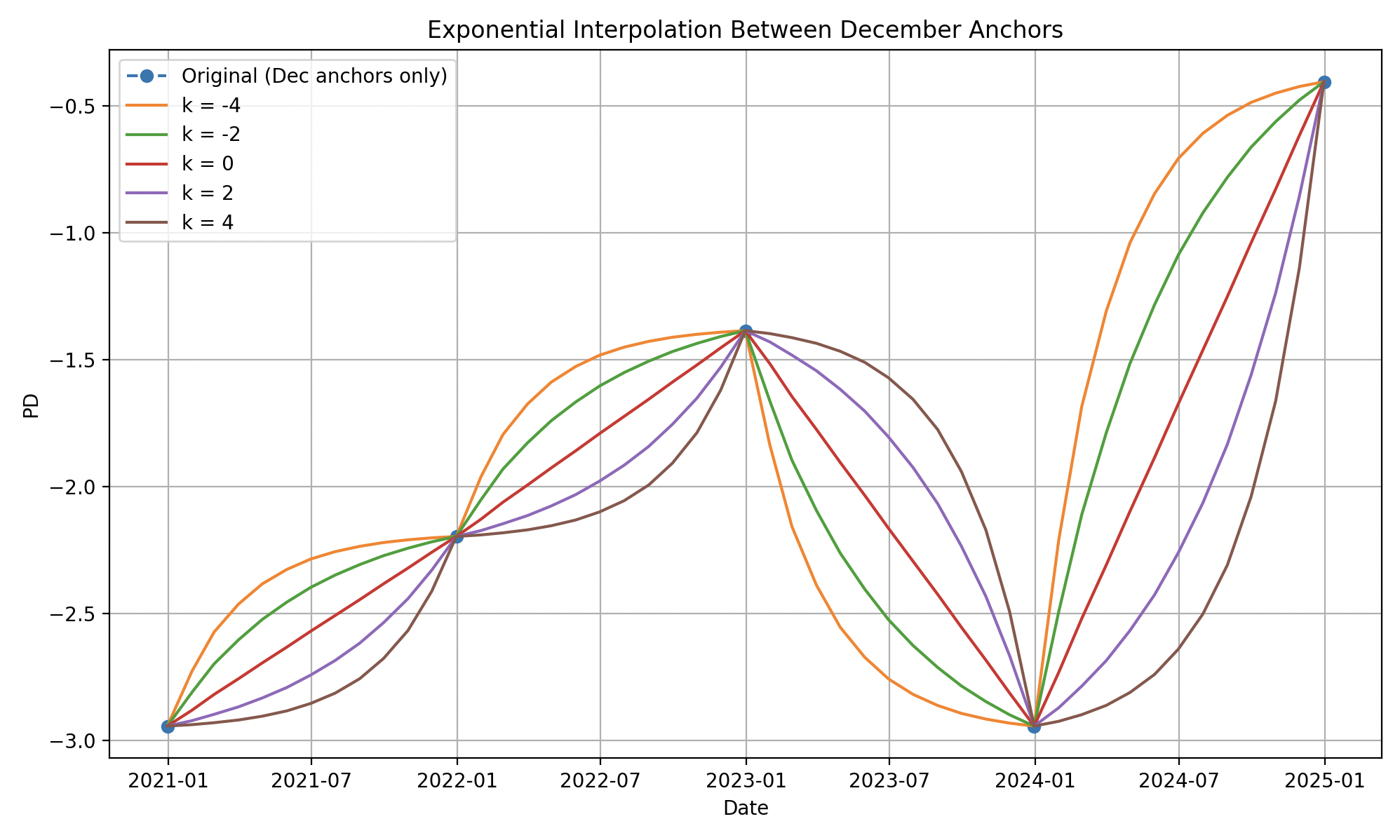}
  \caption{Exponential interpolation with k=3 selected through backtesting}
  \label{fig:k_interpolation}
\end{figure}

The dynamic target is the interpolated static prediction between consecutive December 31st anchor points.
Static model predictions are interpolated exponentially with parameter within the score space.
Exponential interpolation (rather than linear) reflects credit risk dynamics where deterioration typically accelerates as companies approach distress.
Figure \ref{fig:k_interpolation} demonstrates the effect of different k values on the interpolation curve between anchor points.
Positive k values create convex curves with gradual initial transitions and accelerating changes near the next anchor, while negative values produce the opposite pattern. 

\section{Modeling}
\label{section:modeling}

We employ logistic regression as the meta-learner, treating base-model PD outputs as learned representations encoding distinct temporal views of risk.
This modular abstraction preserves interpretability while enabling temporal alignment and integration of new indicators without retraining base models.

\subsection{Static Model Architecture}
\label{subsection:static_model}

Logistic regression trained on December 31st-aligned data combines standardized logit-transformed probabilities with turnover class indicators:

\begin{equation*}
  logit(p) = \beta_0 + \beta_1 \cdot logit(PD_{bhv})^* + \beta_2 \cdot logit(PD_{crd})^* + \sum_{k \in \{\text{micro, small, medium}\}} \gamma_k \cdot I_{turnover,k}
\end{equation*}

where $p$ is the predicted 12-month default probability, and large company size serves as the reference level.

Table \ref{tab:static_coefficients} reports the estimated coefficients.

\begin{table}[ht!]
  \centering
  \caption{Static Model Logistic Regression Coefficients}
  \label{tab:static_coefficients}
  \begin{tabular}{lcccc}
    \hline
    \textbf{Variable} & \textbf{Coefficient} & \textbf{Std. Error} & \textbf{z-value} & \textbf{P-value} \\
    \hline
    Intercept & $-5.1478$ & 0.133 & -38.73 & 0.000 \\
    $logit(PD_{crd})^*$ (crd\_pd) & 0.9770 & 0.042 & 23.28 & 0.000 \\
    $logit(PD_{bhv})^*$ (bhv\_pd) & 0.5839 & 0.029 & 19.95 & 0.000 \\
    C(crd\_size)[medium] & 0.3173 & 0.167 & 1.90 & 0.057 \\
    C(crd\_size)[micro] & 1.1503 & 0.159 & 7.22 & 0.000 \\
    C(crd\_size)[small] & 0.5730 & 0.156 & 3.66 & 0.000 \\
    \hline
    \multicolumn{5}{l}{\textit{Model Statistics: McFadden $R^2$ = 0.338, Nagelkerke $R^2$ = 0.366, KS = 0.652}} \\
    \hline
  \end{tabular}
\end{table}

\subsection{Dynamic Model Architecture}
\label{subsection:dynamic_model}

\begin{figure}[ht!]
  \centering
  \includegraphics[width=0.85\textwidth]{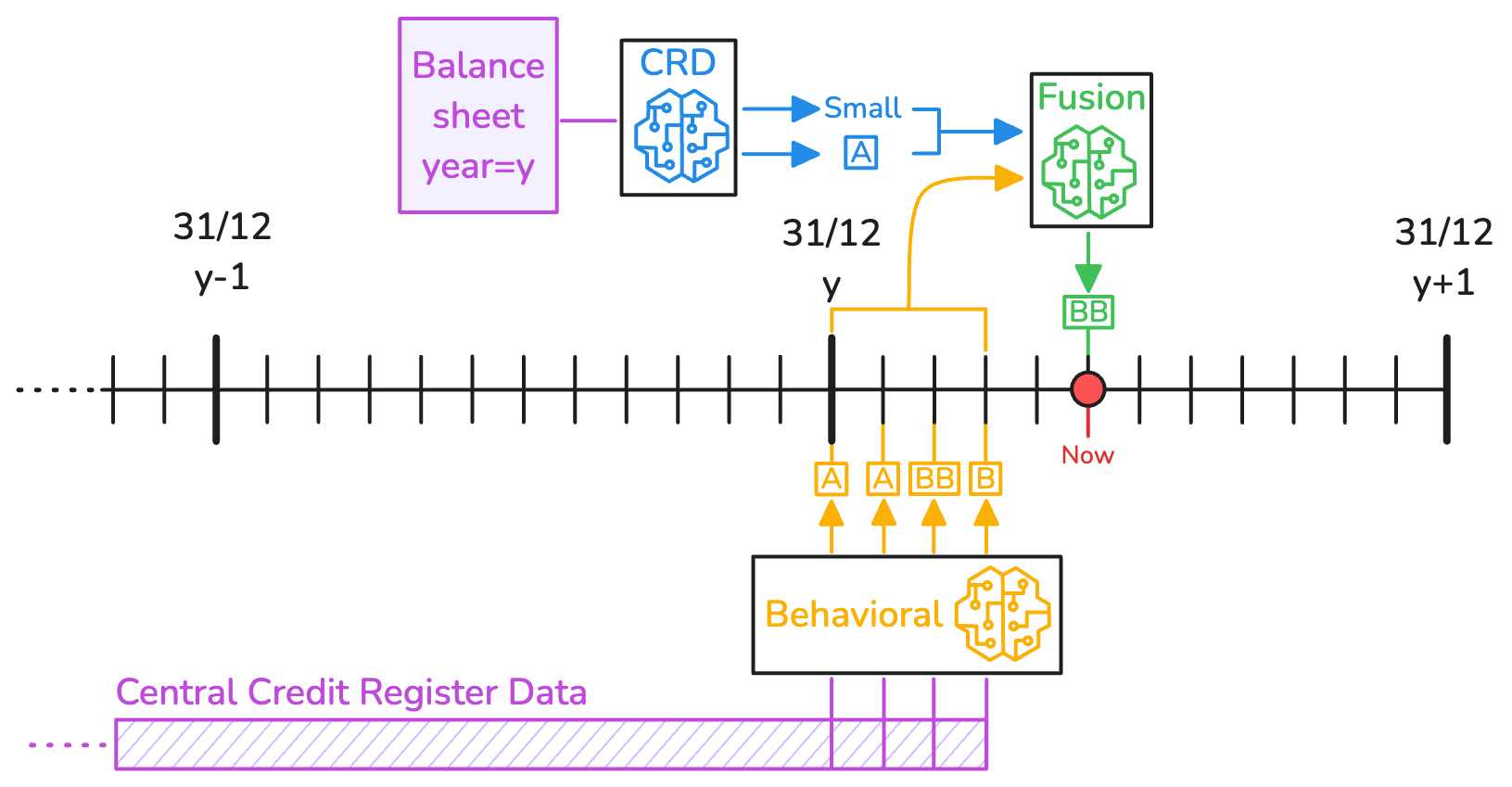}
  \caption{Fusion architecture combining static balance sheet assessment with dynamic behavioral updates at evaluation point}
  \label{fig:dynamic_concept}
\end{figure}

Figure \ref{fig:dynamic_concept} illustrates the temporal alignment concept underlying the dynamic model, showing how CRD model outputs (anchored at December 31st) and behavioral scores are integrated at the evaluation point to produce the fusion prediction.

The dynamic model uses EWMA behavioral scores with the same equation structure as the statics model.

\begin{equation*}
  logit(\hat{p}_{t+k}) = \beta_0 + \beta_1 \cdot logit(PD_{bhv})_k^* + \beta_2 \cdot logit(PD_{crd})^* + \sum_{s \in \{\text{micro, small, medium}\}} \gamma_s \cdot I_{turnover,s}
\end{equation*}

Table \ref{tab:dynamic_coefficients} presents coefficients ($R^2 = 0.952$, 149,353 observations).

\begin{table}[ht!]
  \centering
  \caption{Dynamic Model OLS Regression Coefficients}
  \label{tab:dynamic_coefficients}
  \begin{tabular}{lcccc}
    \hline
    \textbf{Variable} & \textbf{Coefficient} & \textbf{Std. Error} & \textbf{t-value} & \textbf{P-value} \\
    \hline
    Intercept & $-4.9864$ & 0.002 & -2962.3 & 0.000 \\
    $logit(PD_{crd})^*$ (crd\_pd) & 0.9960 & 0.001 & 933.2 & 0.000 \\
    $logit(PD_{bhv})^*$ (bhv\_pd\_ewma) & 0.7913 & 0.001 & 743.5 & 0.000 \\
    C(crd\_size)[medium] & 0.2992 & 0.003 & 118.9 & 0.000 \\
    C(crd\_size)[micro] & 1.1362 & 0.003 & 374.2 & 0.000 \\
    C(crd\_size)[small] & 0.5612 & 0.003 & 221.9 & 0.000 \\
    \hline
  \end{tabular}
\end{table}

\subsection{Rating Assignment and Meta-Learning Integration}
\label{subsection:delta_shift}


The goal of the enriched system is to provide rating classes comparable to established credit rating systems.
Therefore, we map continuous PD predictions to categorical ratings using the CRD rating scale.
To ensure proper calibration across company size segments, we apply size-specific delta shifts that adjust the predicted PDs before rating assignment.
These delta shifts account for systematic differences in default behavior and data quality across size categories.

\begin{table}[h!]
  \centering
  \caption{Delta Shift Values by Company Size}
  \label{tab:delta_shift}
  \begin{tabular}{lc}
    \hline
    \textbf{Size Category} & \textbf{Delta Shift} \\
    \hline
    Micro & -2.004 \\
    Small & -1.251 \\
    Medium & -0.688 \\
    Large & +0.178 \\
    \hline
  \end{tabular}
\end{table}


\subsection{Validation Strategies}
\label{subsection:validation}

To ensure the model's applicability across the entire banking unit and its robustness under different scenarios, we employ multiple validation perspectives.
We cover temporal, cross-sectional, and bootstrap dimensions.
Each validation strategy addresses specific aspects of model performance and generalization

\section{Results}
\label{section:results}

Table \ref{tab:static_performance} demonstrates clear performance gains from temporal alignment. The Static Fusion model achieves AUC=0.900, outperforming CRD (0.833) and BHV (0.808) baselines.

\begin{table}[ht!]
  \centering
  \caption{Static Model Performance Metrics}
  \label{tab:static_performance}
  \begin{tabular}{lcccc}
    \hline
    \textbf{Model} & \textbf{AUC} & \textbf{F-measure} & \textbf{Recall} & \textbf{Specificity} \\
    \hline
    Fusion Static & 0.900 & 0.821 & 0.820 & 0.822 \\
    Behavioral (baseline) & 0.808 & 0.721 & 0.662 & 0.818 \\
    CRD (baseline) & 0.833 & 0.689 & 0.591 & 0.882 \\
    \hline
  \end{tabular}
\end{table}

The model demonstrates robust performance across different company size segments, with particularly strong results for medium and large enterprises as shown in Table \ref{tab:size_performance}.

\begin{table}[ht!]
  \centering
  \caption{Performance Metrics by Company Size}
  \label{tab:size_performance}
  \begin{tabular}{lccccccc}
    \hline
    \textbf{Size} & \textbf{Count} & \textbf{Default Rate} & \textbf{AUC} & \textbf{Avg. Precision} & \textbf{F-measure} & \textbf{Recall} & \textbf{Specificity} \\
    \hline
    micro & 2,275 & 0.0615 & 0.882 & 0.421 & 0.741 & 0.907 & 0.597 \\
    small & 4,781 & 0.0316 & 0.882 & 0.439 & 0.782 & 0.801 & 0.758 \\
    medium & 4,894 & 0.0206 & 0.903 & 0.428 & 0.829 & 0.802 & 0.867 \\
    large & 6,504 & 0.0134 & 0.871 & 0.242 & 0.801 & 0.736 & 0.908 \\
    \hline
  \end{tabular}
\end{table}

The dynamic model maintains strong performance across temporal horizons (Table \ref{tab:dynamic_horizons}), with QWK ranging from 0.772 (6-month ahead) to 0.951 (12-month ahead).

\begin{table}[ht!]
  \centering
  \caption{Dynamic Model Performance Across Time Horizons}
  \label{tab:dynamic_horizons}
  \begin{tabular}{lc}
    \hline
    \textbf{Time Horizon}  & \textbf{Quadratic Weighted Kappa} \\
    \hline
    1-month ahead & 0.910 \\
    3-month ahead  & 0.849 \\
    6-month ahead  & 0.772 \\
    12-month ahead & 0.951 \\
    \hline
  \end{tabular}
\end{table}

\begin{figure}[ht!]
  \centering
  \includegraphics[width=0.9\textwidth]{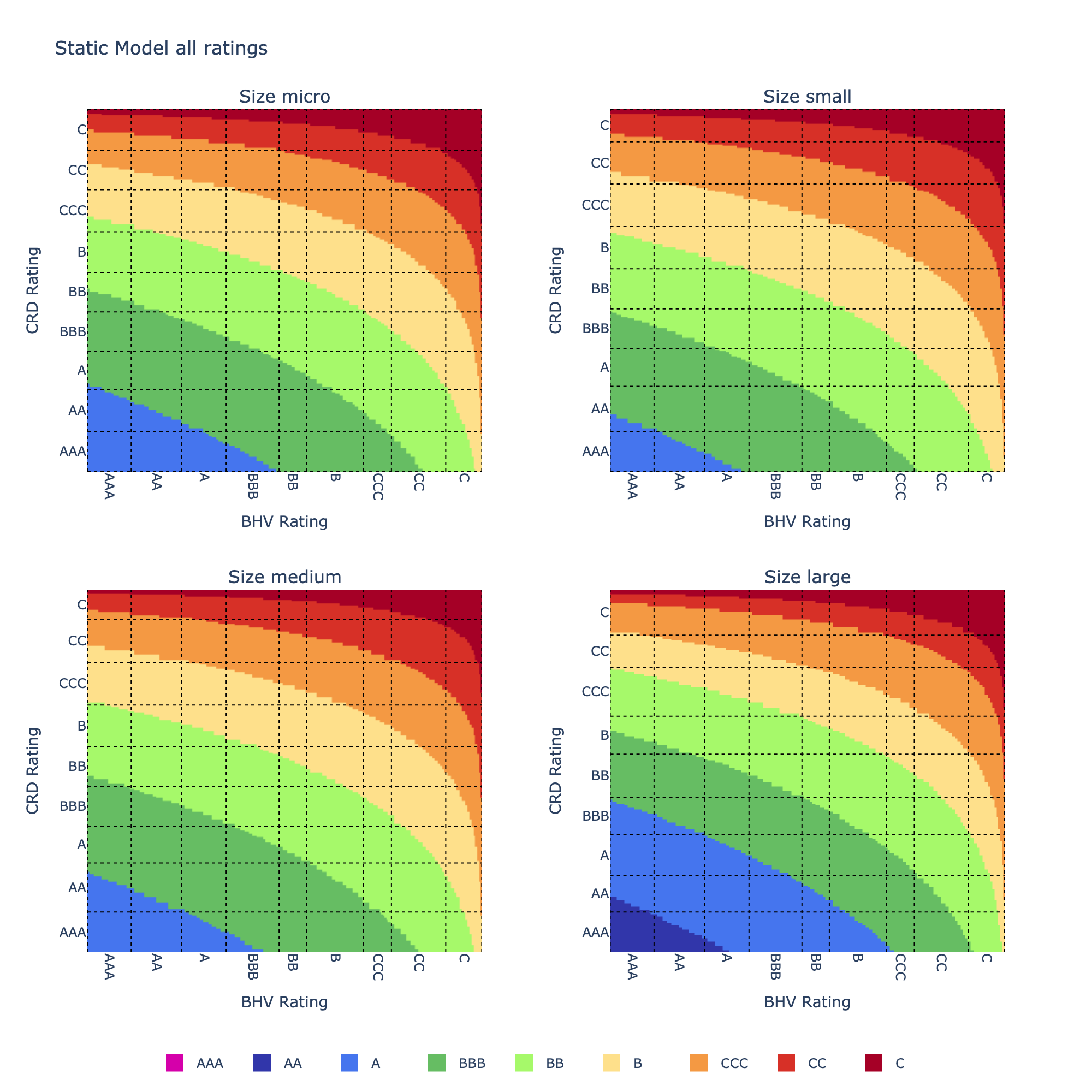}
  \caption{Fusion rating frontier by company size}
  \label{fig:rating_frontier}
\end{figure}

Figure \ref{fig:rating_frontier} illustrates fusion rating integration across all company sizes.

\section{Discussion}
\label{section:discussion}

\paragraph{Limitations}

The dynamic model assumes smooth risk evolution between balance sheet assessments.
While sudden economic shocks may violate this assumption,
high-frequency behavioral data through EWMA captures rapid quality changes.
Balance sheet annual updates create a practical constraint: between publications (potentially 18 months),
only behavioral components drive prediction changes.
The adjusted default definition may not perfectly capture institution-specific defaults,
particularly for complex multi-bank relationships.

\paragraph{Choice of Meta-Learner}

We selected logistic regression over Random Forest \cite{friedman2002stochastic} after several experiments showed no significant AUC or log-loss improvements. The simpler approach preserves modularity and interpretability, aligning with our temporal decomposition contribution.

\paragraph{Generalizability}

While developed for Italian SMEs, the framework applies broadly to markets with heterogeneous data sources and publication delays.
Parameters ($\alpha$, k, delta shifts) are market-specific, but the static-dynamic architectural decomposition generalizes.
Markets with complete data availability (e.g., US) may benefit less; emerging markets could see larger improvements.

\paragraph{Implications for Credit Risk Practice}

Our framework challenges the conventional wisdom that more frequent model updates necessarily improve risk assessment.
By explicitly modeling temporal dynamics rather than repeatedly retraining on misaligned data, we achieve superior performance with lower operational complexity.

\paragraph{Explainability}

The framework provides interpretability through meta-model coefficients (component importance), delta shifts (size-based attribution), and marginal effects.
SHAP analysis \cite{lundberg2017unified} in hierarchical stacking requires careful decomposition; we rely primarily on coefficient interpretation and marginal effects \cite{jean2025bridging}.

\section{Conclusion}
\label{section:conclusion}

This paper demonstrates that temporal misalignment in credit data sources is a fundamental challenge requiring architectural innovation.
By decomposing risk assessment into static anchoring and dynamic evolution, we transform the liability of asynchronous data into an asset—leveraging each source's temporal characteristics.
Obtained evidences validate this approach: explicit temporal modeling yields both statistical improvements and operational advantages in both credit origination and portfolio monitoring.

The three key innovations—temporal decomposition, point-in-time consistency, and modular meta-learning—collectively address a gap that existing multi-view and temporal architectures have overlooked: the operational reality of asynchronous data availability.




\section*{Acknowledgements}

We would like to thank Claudio Nordio, Risk Officer, for supporting this work and research. We are also grateful to Lorenzo Giada for assessing our approach and providing valuable suggestions that improved this paper.

\bibliographystyle{plain}
\bibliography{fusion_paper}

\end{document}